\documentclass[12pt]{iopart}

\def\bra#1{\mathinner{\langle{#1}|}}
\def\ket#1{\mathinner{|{#1}\rangle}}

\newcommand{\mrm}{\mathrm}

\newcommand{\ii}{\mrm{i}}

\newcommand{\pp}{\mrm{p}}
\newcommand{\cc}{\mrm{c}}
\newcommand{\mw}{\mu}
\newcommand{\gp}{\gamma_\mathrm{p}}
\newcommand{\gc}{\gamma_\mathrm{c}}
\newcommand{\gr}{\gamma_\mathrm{rel}}

\newcommand{\ee}{\boldsymbol{\epsilon}}
\newcommand{\bfE}{\textbf{E}}

\usepackage{iopams}
\usepackage{mathbbol}              % for math symbols.
\usepackage{graphics}
\usepackage{graphicx}
\usepackage{epsfig}
\usepackage{ulem}
\usepackage{subfigure}
\usepackage{bbm}
\usepackage{url}
\usepackage{color}

\begin{document}

\title{Microwave dressing of Rydberg dark states}
%\date{\today}
\author{M. Tanasittikosol, J. D. Pritchard, D. Maxwell, A. Gauguet, K. J. Weatherill, R. M. Potvliege, and C. S. Adams}
\address{Department of Physics, Durham University,
Rochester Building, South Road, Durham DH1 3LE, UK}

\begin{abstract}
We study electromagnetically induced transparency (EIT) in the 5s$\rightarrow$5p$\rightarrow$46s ladder system of a cold $^{87}$Rb gas. We show that the resonant microwave coupling between the 46s and 45p states leads to an Autler-Townes splitting of the EIT resonance. This splitting can be employed to vary the group index by $\pm 10^5$ allowing independent control of the propagation of dark state polaritons. We also demonstrate that microwave dressing leads to enhanced interaction effects.
In particular, we present evidence for a
$1/R^3$ energy shift between Rydberg
states resonantly coupled by the microwave field and the ensuing breakdown
of the pair-wise interaction approximation.
\end{abstract}

\pacs{32.80.Rm,42.50.Gy,03.67.Lx}
\maketitle

\section{Introduction}

The application of electromagnetic (EM) fields to control the propagation of light through a medium has widespread applications in non-linear optics \cite{boyd}. One important example is electromagnetically induced transparency (EIT) \cite{eit_review} where an optical control field modifies the transmission \cite{harris} and the group velocity of light \cite{hau1999}. The propagation of light through an EIT medium can be described in terms of dark state polaritons \cite{lukin00}. By varying the control field one can reversibly convert between the light and atomic excitations and thereby implement photon storage \cite{chan05,eisa05}.

The use of highly excited Rydberg states in EIT \cite{moha07} creates additional possibilities due to their extremely strong interparticle interactions and extreme sensitvity to external electric fields \cite{gallagher,comp2010}. For example, Rydberg EIT has been used to modify the frequency of light \cite{moha08}, study interactions in cold Rydberg gases \cite{weat08,rait09,prit2010} and probe electric fields inside vapour cells \cite{baso2010,kueb10} and close to surfaces \cite{amsterdam}. A feature of Rydberg states is the large dipole moment to nearby states which scales with the square of the principal quantum number, $n^2$. The large dipole moments are exploited to achieve the strong coupling regime in cavity QED \cite{remp90,raim01}. Transitions between neighbouring Rydberg states are typically in the microwave or millimeter wave regime. Microwave or millimeter wave coupling is of interest in the context of precision measurement of the Rydberg quantum defects \cite{li03}, electric fields \cite{ost99}, and studies of multiphoton ionization \cite{maho91}, which can exhibit features of dynamical localisation \cite{blum91,sche2009}. In addition the microwave field can be used to enhance resonant energy transfer to dipole-coupled pair-states \cite{bohlouli07}.

In this work we study the effect of resonant microwave fields on EIT involving highly excited Rydberg states. By modelling the experimental data we show that adding microwave coupling between Rydberg states can switch the group index of the sample by $\pm 10^5$. We also show that microwave dressing leads to complete or almost complete suppression of the EIT due to enhanced interaction effects.

\section{Experiment}

\begin{figure}[]
\begin{center}
\includegraphics[width=8.6cm]{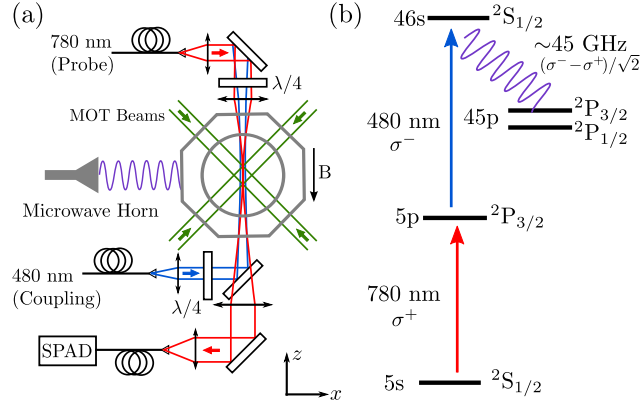}
\caption[]{(a) Schematic of the experimental setup. The probe and coupling beams counter-propagate through a cold Rb cloud. Microwaves are applied from a perpendicular direction. The probe transmission is measured using a single-photon avalanche detector (SPAD). (b) Simplified level scheme showing microwave coupling between the 46S$_{1/2}$ and 45P$_{1/2,3/2}$ Rydberg states.\label{fig1}}
\end{center}
\end{figure}

The experiments are performed on a cloud of laser-cooled $^{87}$Rb atoms using the experimental setup described in \cite{prit2010}. The experimental setup and atomic level scheme are shown schematically in \fref{fig1} (a) and (b) respectively. The atoms are loaded into a magneto-optical trap (MOT) in 1~s and then after optical molasses are prepared in the 5s~$^2$S$_{1/2}$ $|F=2, m_F=2\rangle$ state by optical pumping. EIT spectroscopy is then performed using counter-propagating probe and coupling laser beams focused to 1/e$^2$ radii of 12~$\mu$m and 66 $\mu$m, respectively. The coupling laser beam is stabilised to the 5p $^2$P$_{3/2}$ $(F' = 4)$ $\rightarrow$ $n$s $^2$S$_{1/2}$ $(F'')$ transition using an EIT locking scheme \cite{abel09}.
The probe beam is scanned over the 5s $^2$S$_{1/2}(F = 2)$ $\rightarrow$ 5p~$^2$P$_{3/2}$ $(F' = 3)$ transition in 500~$\mu$s using an acousto-optic modulator. Probe transmission is recorded using a single photon avalanche photodiode (SPAD), averaging over 100 repeats for each dataset. A microwave source (Anritsu MG3696A) applies a field from a direction orthogonal to the probe laser axis using a waveguide. The plane of polarisation of the microwave field is orthogonal to the direction of the pump and probe beams. This configuration leads to microwave coupling to multiple magnetic sub-levels (see section 3) and was constrained by the experimental geometry. The microwave transition frequencies between Rydberg states were calculated using quantum defects from Li {\it et al.} \cite{li03} and are 44.559 and 43.415 GHz for 46S$_{1/2}$ $\rightarrow$ 45P$_{1/2,3/2}$, respectively. EIT spectra were recorded for varying microwave power.

\begin{figure}[t!]
\begin{center}
\includegraphics[trim=0cm 0cm 0cm 0.cm, clip=true,width=8.0cm]{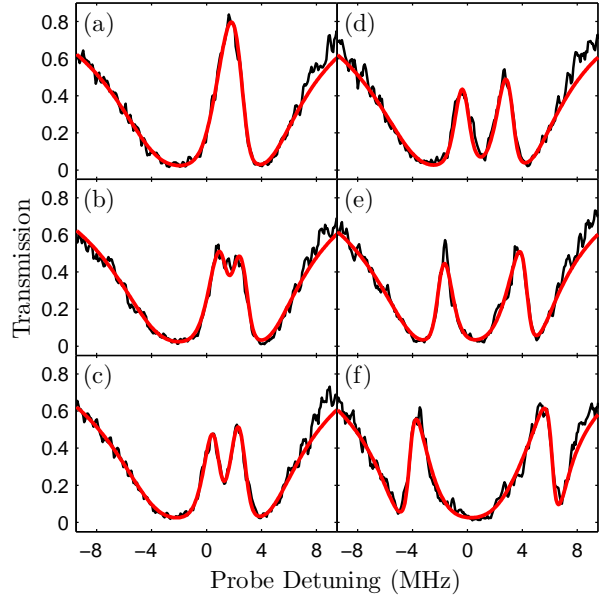}
\caption[]{EIT spectra with increasing microwave coupling. The microwave Rabi frequencies, $\Omega_{\mu}^r$ from the fit parameters, are (a) 0, (b) 2.2, (c) 3.6, (d) 6.8 (e) 12.2, and (f) $21.0 \times 2\pi$ MHz; these values match the scaling of microwave power in the experiment, although the microwave electric field cannot be measured. \label{fig2}}
\end{center}
\end{figure}

Figure \ref{fig2} shows the evolution of the EIT signal with increasing microwave power. As the strength of the microwave coupling is increased the EIT peak undergoes an Autler-Townes splitting due to the dressing of the Rydberg state. To understand the evolution of the spectra with increasing microwave power we fit the data using the model described in section 3. The fits obtained from the model are also shown in \fref{fig2}.

\section{Theoretical Modelling}

\begin{figure}[h]
\begin{center}
\includegraphics[width=8.6cm]{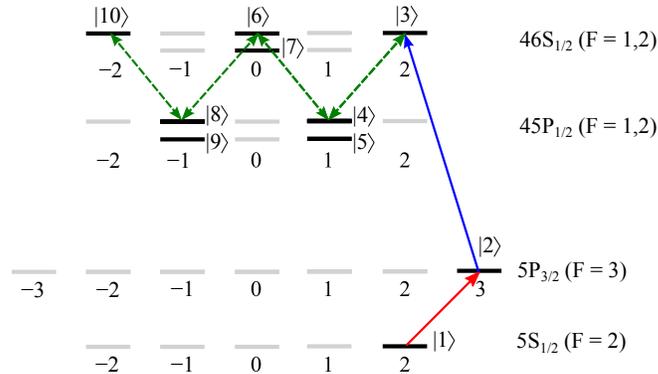}
\caption[]{Schematic of the level scheme used to model the system. \label{fig3}}
\end{center}
\end{figure}

The experiment is modelled using the 10-level atom shown in \fref{fig3} interacting with an EM field given by
\begin{equation}\label{eq:totalfield}
\bfE(t)={1\over\sqrt{2}}E_{\pp}\hat{\ee}_{+}\rme^{-\ii\omega_{\pp}t}+{1\over\sqrt{2}}E_{\cc}\hat{\ee}_{-}\rme^{-\ii\omega_{\cc}t}+{1\over2}E_{\mw}\hat{{\bf x}}\rme^{-\ii\omega_{\mw}t}+{\rm c.c.}~,
\end{equation}
where $\hat{\ee}_{\pm}$ are the polarization unit vectors representing $\sigma^{\pm}$ transitions and $\hat{{\bf x}}$ is the polarization unit vector in the $x$ direction. The first term in \eref{eq:totalfield} represents the probe field, whose amplitude and frequency are $E_{\pp}$ and $\omega_{\pp}$, respectively. The second term represents the coupling field with amplitude $E_{\cc}$ and frequency $\omega_{\cc}$. The third term is the microwave field with amplitude $E_{\mw}$ and frequency $\omega_{\mw}$. Using a quantization axis along $z$, the linearly-polarized microwave field is described as the superposition of $\hat{\ee}_{\pm}$, i.e. $\hat{{\bf x}}=(\hat{\ee}_{-}-\hat{\ee}_{+})/\sqrt{2}$,
% Hence, the states  $m_{F}=0$ of $F=1$ (labeled as $\ket{7}$) and $m_F=-2,0,2$ of $F=2$ (labeled as $\ket{10}$, $\ket{6}$ and $\ket{3}$, respectively) of 46S$_{1/2}$ couple to the states $m_{F}=-1,1$ of $F=1$ (labeled as $\ket{5}$ and $\ket{9}$, respectively) and $m_F=-1,1$ of $F=2$ (labeled as $\ket{8}$ and $\ket{4}$, respectively) of 45P$_{1/2}$ via the $\sigma^{\pm}$ transitions of the microwave field.
leading to the W-shaped coupling between the Rydberg states (dotted (green) lines in \fref{fig3}). The hyperfine splitting between the $F=1$ and 2 Rydberg states is neglected, i.e., states $|3\rangle, |6\rangle,|7\rangle$ and $|10\rangle,$ and, $|4\rangle,|5\rangle,|8\rangle$ and $|9\rangle$, are assumed to be degenerate. This is justified as the Rabi frequency of the microwave transition is significantly larger than the hyperfine splitting of the Rydberg levels (about $400 \times 2\pi$ kHz \cite{li03}).

Applying the rotating-wave approximation and the slowly-varying variables transformation, the Hamiltonian of the system is given by $\mathcal{H}=\mathcal{H}_0+\mathcal{H}_\mrm{EIT}+\mathcal{H}_\mw$, where
\numparts
\begin{eqnarray} \label{eq:Hamiltonian}
{\scriptsize\fl \mathcal{H}_0\!\!=\!\!-\!\Delta_1\!\ket{2}\!\!\bra{2}\!-\!\Delta_2(\ket{3}\!\!\bra{3}\!+\!\ket{6}\!\!\bra{6}\!+\!\ket{7}\!\!\bra{7}\!+\!\ket{10}\!\!\bra{10})\!-\!\Delta_3(\ket{4}\!\!\bra{4}\!+\!\ket{5}\!\!\bra{5}\!+\!\ket{8}\!\!\bra{8}\!+\!\ket{9}\!\!\bra{9})}, \\
{\scriptsize\fl \mathcal{H}_\mrm{EIT} = \frac{\hbar\Omega_\pp}{2}\ket{1}\!\!\bra{2}+\frac{\hbar\Omega_\cc}{2}\ket{2}\!\!\bra{3}+(\mathrm{h.c.}),}\\
{\scriptsize\fl \mathcal{H}_\mw = \!\!\!\!\!\!\displaystyle\sum_{i=\{3,5,7,9\}}\!\!\!\!\!\!\frac{\hbar\Omega^{(i,i+1)}_\mw}{2}\ket{i}\!\!\bra{i\!+\!1}+\displaystyle\sum_{i=3}^8\frac{\displaystyle \hbar\Omega^{(i,i+2)}_\mw}{\displaystyle 2}\ket{i}\!\!\bra{i\!+\!2}+\!\!\!\!\displaystyle\sum_{i=\{4,6\}}\!\!\!\frac{\hbar\Omega^{(i,i+3)}_\mw}{2}\ket{i}\!\!\bra{i\!+\!3}+(\mathrm{h.c.}).}
\end{eqnarray}
\endnumparts
%\begin{equation}\label{eq:Hamiltonian}
%{\scriptsize \fl{\mathcal{H}}=\hbar\left( \begin{array}{cccccccccc}
%0 & \Omega_{\pp}/2 & 0 & 0 & 0 & 0 & 0 & 0 & 0 & 0\\
%\Omega_{\pp}/2 & -\Delta_{1} & \Omega_{\cc}/2 & 0 &0 &0& 0 & 0 & 0 & 0\\
%0 & \Omega_{\cc}/2 & -\Delta_{2} & 0 & \Omega_{\mw}^{(3,4)}/2 & \Omega_{\mw}^{(3,5)}/2 & 0 & 0 & 0 & 0\\
%0 & 0 & \Omega_{\mw}^{(4,3)}/2& -\Delta_{3} &0& \Omega_{\mw}^{(4,6)}/2 & \Omega_{\mw}^{(4,7)}/2 &0&0&0\\
%0 & 0 & \Omega_{\mw}^{(5,3)}/2 & 0& -\Delta_{3} &\Omega_{\mw}^{(5,6)}/2&\Omega_{\mw}^{(5,7)}/2&0&0&0 \\
%0 & 0 & 0 & \Omega_{\mw}^{(6,4)}/2& \Omega_{\mw}^{(6,5)}/2&-\Delta_{2}&0 &\Omega_{\mw}^{(6,8)}/2&\Omega_{\mw}^{(6,9)}/2&0 \\
%0 & 0 & 0 & \Omega_{\mw}^{(7,4)}/2& \Omega_{\mw}^{(7,5)}/2&0&-\Delta_{2} &\Omega_{\mw}^{(7,8)}/2&\Omega_{\mw}^{(7,9)}/2&0 \\
%0 & 0 & 0 & 0& 0&\Omega_{\mw}^{(8,6)}/2&\Omega_{\mw}^{(8,7)}/2&-\Delta_{3} &0&\Omega_{\mw}^{(8,10)}/2 \\
%0 & 0 & 0 & 0& 0&\Omega_{\mw}^{(9,6)}/2&\Omega_{\mw}^{(9,7)}/2&0&-\Delta_{3} &\Omega_{\mw}^{(9,10)}/2 \\
%0 & 0 & 0 & 0& 0&0&0&\Omega_{\mw}^{(10,8)}/2&\Omega_{\mw}^{(10,9)}/2&-\Delta_2
%\end{array} \right),}
%\end{equation}
Here $\Delta_1\equiv\hbar\Delta_{\pp}$, $\Delta_2\equiv\hbar(\Delta_\pp+\Delta_\cc)$, $\Delta_3\equiv\hbar(\Delta_\pp+\Delta_\cc-\Delta_\mw)$ and, $\Delta_\pp$, $\Delta_\cc$ and $\Delta_\mw$ are the detunings of the probe, coupling and microwave fields, respectively. The Rabi frequencies associated with the probe, coupling and microwave fields, $\Omega_\pp$, $\Omega_\cc$ and $\Omega_{\mw}^{(n,m)}$ respectively, are given by
\begin{eqnarray}
\Omega_\pp&=&{\sqrt{2}E_\pp\over\hbar}\bra{2}e{\bf r}\cdot\hat{\ee}_{+}\ket{1},\\
\Omega_\cc&=&{\sqrt{2}E_\cc\over\hbar}\bra{3}e{\bf r}\cdot\hat{\ee}_{-}\ket{2},\\
\Omega_\mw^{(i,j)}&=&{E_\mw\over\sqrt{2}\hbar}(\bra{i}e{\bf r}\cdot\hat{\ee}_{-}\ket{j}-\bra{i}e{\bf r}\cdot\hat{\ee}_{+}\ket{j}),
\end{eqnarray}
where $e \mathbf{r}$ is the dipole operator, and, $i$ and $j$ correspond to the magnetic sublevels of 46S$_{1/2}$ and 45P$_{1/2}$. Using Wigner-Eckart theorem, the Rabi frequency of the microwave field reduces to
\begin{eqnarray}\label{reducedRF}
\fl\Omega_\mw^{(i,j)}=\Omega_\mw^r\times(-1)^{m_{F}^{i}}\sqrt{(2F^{i}+1)(2F^{j}+1)}\left\{ \begin{array}{ccc}J^{i} & J^{j} & 1\\F^{j} & F^{i} & 3/2\end{array}\right\}\left\{ \begin{array}{ccc}L^{i} & L^{j} & 1\\J^{j} & J^{i} & 1/2\end{array}\right\} \nonumber\\
\times\left( \begin{array}{ccc}L^{j} & 1 & L^{i}\\0 & 0 & 0\end{array}\right)\left[\left( \begin{array}{ccc}F^{j} & 1 & F^{i}\\m_{F}^{j} & -1 & -m_{F}^{i}\end{array}\right)-\left( \begin{array}{ccc}F^{j} & 1 & F^{i}\\m_{F}^{j} & 1 & -m_{F}^{i}\end{array}\right)\right],
\end{eqnarray}
where $\Omega_\mw^r = \sqrt{6}E_\mw/\hbar\times\langle 46S_{1/2}\vert er \vert45P_{1/2}\rangle$ contains the radial matrix element calculated as $\langle 46S_{1/2}\vert er \vert45P_{1/2}\rangle=1924$~$e$a$_0$ using the Numerov method \cite{zimm79}.

The equation of motion for the density matrix  $\rho$ of the 10-level system is given by
\begin{equation}\label{eq:Liouville}
{\partial \rho\over\partial t}=-\frac{\ii}{\hbar}[\mathcal{H},\rho]+\mathcal{L}(\rho)+\mathcal{L}_d(\rho),
\end{equation}
where $\mathcal{L}(\rho)=\sum_ic_i\rho c_i^{\dagger}-(c_i^{\dagger}c_i\rho+\rho c_i^{\dagger}c_i)/2$ is the Lindblad superoperator \cite{lindblad76} describing spontaneous decay and $\mathcal{L}_d(\rho)$ is a dephasing matrix which accounts for the linewidth of the EM fields.
The natural decay linewidths of the 46$S_{1/2}$ and 45$P_{1/2}$ states are approximately 2~kHz and can be neglected, so only the decay channel from 5$P_{3/2}$ to 5$S_{1/2}$ at a rate $\Gamma/2\pi = 6$ MHz is included using operator $c=\sqrt{\Gamma}\ket{1}\!\!\bra{2}$. In addition to spontaneous emission, the dephasing due to the finite linewidth of the probe and coupling fields (giving rise to dephasing rates $\gp$ and $\gc$, respectively) is included, as well as the dephasing of the Rydberg states with respect to the other states (rate $\gamma_{{\rm Ry}}$). The latter is most likely due to fluctuating electric and magnetic stray fields. The linewidth of the microwave source is negligible. For EIT the important linewidth is the relative linewidth $\gr$ of the two-photon transition between the probe and coupling laser, typically taken equal to $\gp+\gc$ \cite{gea95}. However, for the EIT locking scheme used to stabilise the coupling laser transition \cite{abel09}, $\gr$ is actually less than the linewidth of either laser. The resulting dephasing matrix $\mathcal{L}_d(\rho)$ is detailed in \ref{sec:L}.

The steady state solution of \eref{eq:Liouville} is found by setting $\partial \rho/\partial t=0$. Using semiclassical theory, the susceptibility, $\chi(\Delta_\pp)$, of the system is proportional to the steady state coherence, $\rho_{21}$, between the intermediate and ground states, i.e.\cite{loudon97},
\begin{equation}
\chi(\Delta_\pp)=-{2Nd_{21}^2\over\hbar\epsilon_0\Omega_{{\rm p}}}\rho_{21}~,
\end{equation}
where $N$ is the atomic density and $d_{21}=\bra{2}e{\bf r}\cdot\hat{\ee}_{+}\ket{1}=1/\sqrt{3}\times5.177$~$e$a$_0$ \cite{siddons08} is the dipole matrix element for the probe transition. The transmission through the medium is then given by the Beer-Lambert law
\begin{equation}\label{eq:transmission}
T=\exp\left({2NLd_{21}^2k_{{\rm p}}\over\hbar\epsilon_0\Omega_{{\rm p}}}{\rm Im}[\rho_{21}]\right)~,
\end{equation}
where $L$ is the length of the atomic cloud and $k_{{\rm p}}=2\pi/\lambda_{{\rm p}}$ is the wavenumber of probe laser. At the relatively low probe powers considered in this work, $\rho_{21}$ is independent of $\gc$; instead it is only through $\gr$ that the linewidth of the coupling laser enters. Setting the column density, CD$\equiv NL$ and assuming the weak probe limit, the transmission becomes a function of 8 parameters, i.e. $\Omega_\cc,\Omega_\mw^{r},\Delta_\cc,\Delta_\mw,\gp,\gr,\gamma_{{\rm Ry}},$ and ${\rm CD}$ which can be determined from fitting experimental data. We began by fitting \eref{eq:transmission} to data using the probe laser only, reducing the system to 2-levels to obtain CD and $\gp/2\pi$ (1.5$\times10^{13}$ m$^{-2}$ and $0.33$ MHz, respectively).
% and their values are .
Subsequently, we fit the 3-level EIT transmission, which determines the quantities $\Omega_\cc/2\pi$, $\Delta_\cc/2\pi$ and $\gr/2\pi$ ($5.5$ MHz, $-1.9$ MHz and $0.14$ MHz, respectively). Finally the remaining three variables, related to the microwave dressing, $\Omega_{\mw}^{r}$, $\Delta_\mw$ and $\gamma_{{\rm Ry}}$ are determined using the 10-level model. $\Omega_{\mu}^r$ scales proportional to the applied microwave field as expected, $\Delta_\mw/2\pi$ fluctuates between $-0.2$ and 0 MHz and $\gamma_{{\rm Ry}}/2\pi$ is 0.3 MHz. Using this method we obtain excellent agreement between the theoretical prediction (red solid curve) and the experimental data (black solid curve) for each microwave power, as shown in \fref{fig2}. The calculated lineshape is sensitive to the number of levels included in the model. Reducing the number of states removes the symmetry in the system, leading to anomalous resonances which are not observed in the experiment. This 10-level W model represents the minimum number of states required to accurately reproduce both the detuning of the microwave splitting and the peak amplitudes.

\section{Group Index}

One attractive aspect of EIT is the possibility to obtain a very high group index resulting in slow light \cite{hau1999}. By varying the group index one can change the mixing angle between the light and matter components of dark state polaritons \cite{lukin00} and thereby implement photon storage \cite{chan05,eisa05}. The group index of the system is given by
\begin{equation}
n_{\mrm{gr}}=n_{\mrm{ph}}+\omega_\pp{\partial n_{\mrm{ph}}\over\partial \omega_\pp}~,
\end{equation}
where $\omega_\pp$ is the frequency of the probe laser and $n_{\mrm{ph}}=1+\mrm{Re}[\chi]/2$ is the refractive index.
The large group index arises from the rapid variation of $n_{\mrm{ph}}$ with $\omega_\pp$ due to the coupling laser. An interesting feature of microwave dressing is the ability to modify the dispersion and hence the dynamics of the Rydberg dark state polaritons \cite{bason08} on relatively fast time scales. In practice, the time response of the dark-states which produce the dispersive feature is limited by EIT transients which depend on the multi-photon Rabi frequency \cite{moha08}.

\begin{figure}[h]
\begin{center}
\includegraphics[trim=0cm 0cm 0cm 0.cm, clip=true,width=8.5cm]{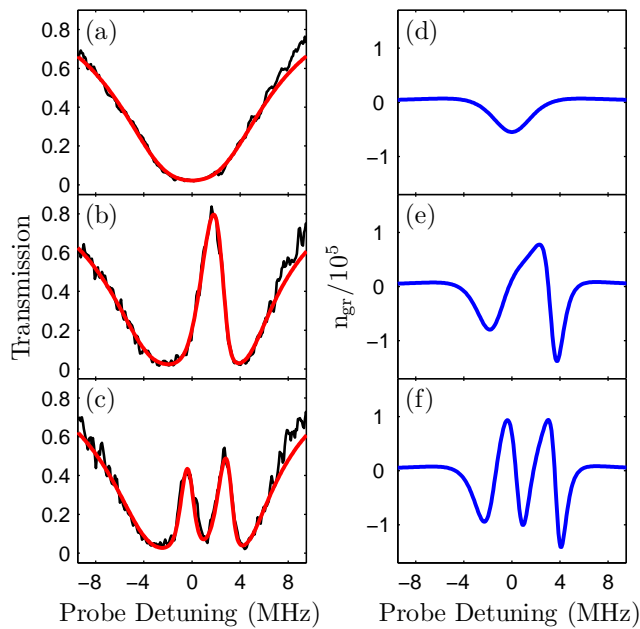}
\caption[]{Transmission and calculated group index, $n_{\rm gr}$ for two-level absorption (a,d), EIT (b,e) and EIT with microwave dressing (c,f). This illustrates how the coupling and microwave fields can be used to control the transparency and pulse propagation speed of the medium.}\label{fig4} \end{center}
\end{figure}

To illustrate the potential of microwave dressing to modify the dispersion we extract the real part of the susceptibility from the 10-level model and use this to calculate the group index which is plotted in \fref{fig4}. We see that on resonance, the microwave field allows independent control of the group index and absorption which could prove useful in controlling the interaction between dark state polaritons. At a probe detuning of $\Delta/2\pi$ = 1~MHz without and with the microwave field the group index is switched from approximately $+5\times10^4$ (figure 4(e)) to $-10^5$ (figure 4(f)) within the transparency window. The negative group index corresponds to ``superluminal'' or backwards propagation \cite{gehring06}, albeit with increased dissipation. However, in contrast to the simple probe-only case \fref{fig4}(a), with microwave dressing one can vary both the microwave and coupling laser powers to trade-off between pulse speed, bandwidth and transparency.

\section{Enhanced Interaction Effects}

All of the data presented above are taken in the weak probe regime of EIT where $\Omega_\mathrm{p}\ll\Omega_\mathrm{c},\Omega_\mw$, resulting in a resonant dark state with all the population in the ground state. As the probe Rabi frequency is increased, dipole-dipole interactions between Rydberg atoms prevent more than a single Rydberg excitation if the resulting energy shift of the Rydberg state,
 $\Delta E(R)$, is larger than the energy width of the two-photon resonance, $\hbar\gamma_\mathrm{EIT}$. This is known as dipole-blockade \cite{lukin01}. The blockade mechanism modifies the EIT dark state as now only a single photon can create transparency in the blockaded volume. This causes a suppression of transmission on resonance \cite{prit2010}. Figures 5(a) and (b) show EIT data taken for a weak and strong probe Rabi frequency, showing the interaction induced suppression. If a weak microwave coupling ($\Omega_\mw<\Omega_\mathrm{c}$) is now introduced from 46S$_{1/2}$ to 45P$_{1/2}$, the suppression is dramatically enhanced, as illustrated in Figures 5(c) and (d).

As the microwave coupling is weak the effect on the group index is small and increased EIT suppression occurs predominantly due to the change in the form of the interactions. Without the microwave field, $\Delta E(R)/\hbar$ for two atoms both in the $46S_{1/2}$ state scales as $1/R^6$ with a
coefficient of $-5.6 \times 2 \pi$  GHz $\mu$m$^6$ \cite{sing05}, giving a blockade radius of approximately 3.5~$\mu$m for $\gamma_\mathrm{EIT}/2\pi=3$~MHz. The microwaves however drive a resonant coupling, resulting in an interaction shift scaling as $1/R^3$. (The coefficient is approximately 
$-0.8 \times 2 \pi$ GHz $\mu$m$^3$ using the dipole matrix element from above.) As illustrated in figure 5(e), this $1/R^3$ potential can be expected to yield a larger blockade radius, hence a larger average number of blockaded atoms and an enhanced suppression of EIT. For two atoms prepared in the superposition
of 
Rydberg states driven by the resonant microwave field, the
blockade radius would be approximately 7~$\mu$m, neglecting any
interaction with Rydberg atoms outside the blockade sphere. 

To evaluate the role of the van der Waals interaction in the change in the
transmission between (a) and (b), we develop the following simple model.
We postulate that
any non-blockaded atom driven by the two laser fields 
has a probability $p_{\rm bl}$ to be in the dark state and 
inhibit the formation of a dark state in any atom located at a distance $R \leq 
R_{\rm bl}$,
a probability $1-p_{\rm bl}$ to be in a dark state and not inhibit the formation of a dark state
in neighbouring atoms, and
a zero probability of affecting atoms beyond $R_{\rm bl}$.
We define the blockade radius $R_{\rm bl}$ by the equation 
$|\Delta E(R_{\rm bl})/\hbar| = \gamma_{\rm EIT}$, where $\Delta E(R)$ is the $1/R^6$ shift shown
in Figure 5(e), and we set $p_{\rm bl}=\rho_{33}$ with $\rho_{33}$ the
population of the $46S_{1/2}$ state calculated as described in Section 3.
($\rho_{33}$ does not exceed about 0.3 for the parameters considered.)
We thus assume that
any blockaded atom 
scatters the probe laser as if the coupling laser was not
present and that any non-blockaded atom forms
a transparent dark state as if the 
Rydberg-Rydberg interaction was not present. Accordingly, 
we take
the susceptibility to be
$\chi=
({N}_{\mrm d} \chi_{\mrm{d}}+
{N}_{\rm bl}\chi_{2})/({N}_{\mrm d}+{N}_{\rm bl})$ where $\chi_{\mrm{d}}$ is the susceptibility of an atom in the dark state,
$\chi_{2}$
the susceptibility of a two-level atom, and ${N}_{\mrm d}$ and ${N}_{\rm bl}$ are the number densities
of dark state and blockaded atoms, respectively. 
Assuming that the
blockade spheres do not overlap, we can write
\begin{equation}
\chi= {\chi_{\mrm{d}} + p_{\rm bl} \mathcal{N} \chi_{2} \over 1 +  p_{\rm bl} \mathcal{N}},
\label{eq:chiav}
\end{equation}
where
$\mathcal{N}=4\pi NR_{\rm bl}^3/3$. ($N$ is the local atomic density, which we
derive from the experimental CD assuming a Gaussian density profile. 
For a constant $N$, a blockade sphere contains $\mathcal{N}$ blockaded atoms on average,
since the atoms are distributed randomly and the centre of the sphere, which is occupied by a dark state atom, is of zero measure.
Typically, $\mathcal{N} \approx 5$ for $R_{\rm bl}= 3.5$ $\mu$m.)

The weak-probe data in (a) and (c) is fitted using the procedure described above to obtain the model parameters shown in \tref{tab:tab2}. The local susceptibilities $\chi_{\mrm{D}}$, $\chi_{2}$ and $\chi$ are then calculated using the same parameters but with the higher probe Rabi frequency of figure 5(b), 
taking into account the spatial intensity profile of the probe beam (including
its attenuation as it passes through the medium). The resulting transmission
profile of the atomic cloud is represented by a solid curve in figure 5(b). 
For the parameters considered,
the transmission profile obtained by correcting equation (\ref{eq:chiav}) for 
the overlap of the different blockade spheres is almost the
same as without correction, except for $\Delta_{\rm p} \approx 0$ where it is up to 4\% higher. 
We conclude from the reasonable agreement between the model and the data that the decrease in the experimental
transmission between
(a) and (c) is consistent with a blockade of Rydberg excitation by
the van der Waals interaction. 

Using the same model to calculate the transmission with the microwave field
present, with $p_{\rm bl}=\rho_{33}$ and the blockade radius still given by the 
$46S_{1/2}-46S_{1/2}$ van der Waals interaction,
leads to the result represented by the solid curve in figure 5(d).
This result is in clear disagreement with the experimental transmission.
(The $45P_{1/2}-45P_{1/2}$ van der Waals interaction is unimportant as far as
determining the blockade radius is concerned since the $45P_{1/2}$ is not
directly coupled to the $5P_{3/2}$ state.)
A better match to the data is obtained by
taking $p_{\rm bl}$ to be the total population in the Rydberg states and
$\Delta E(R)$ to be the $1/R^3$ dipole-dipole shift
when determining the blockade radius (the dotted curve). However, the model still underestimates
the suppression of EIT, and the data is more
closely approximated 
by the transmission
profile calculated assuming that the
whole atomic cloud is blockaded (the dashed curve).
These results
indicate that the
$1/R^3$ dipole-dipole interaction
between Rydberg states resonantly coupled by the microwave field is significant and point to
the ensuing breakdown of the approximation of a pair-wise interaction \cite{amthor09}.

\begin{table}[!h]
\caption{\label{tab:tab2}The values of the free parameters which produce the minimum residuals in fitting the data of Figures 5 (a) and (c). Apart from CD, these parameters are expressed in units of $2\pi$ MHz}.
\begin{indented}
\item[]\begin{tabular}{@{}llllllll}
\br
$\Omega_\cc$&$\Omega^r_\mw$&$\Delta_\cc$&$\Delta_\mw$&$\gp$&$\gr$&$\gamma_{{\rm Ry}}$&${\rm CD}$~(m$^{-2}$)\\
\mr
5.50&2.86&0.64&$-$1.09&0.33&0.08&0.36&1.40$\times10^{13}$\\
\br
\end{tabular}
\end{indented}
\end{table}

\begin{figure}[h]
\begin{center}
\includegraphics[trim=0cm 0cm 0cm 0.cm, clip=true,width=8.6cm]{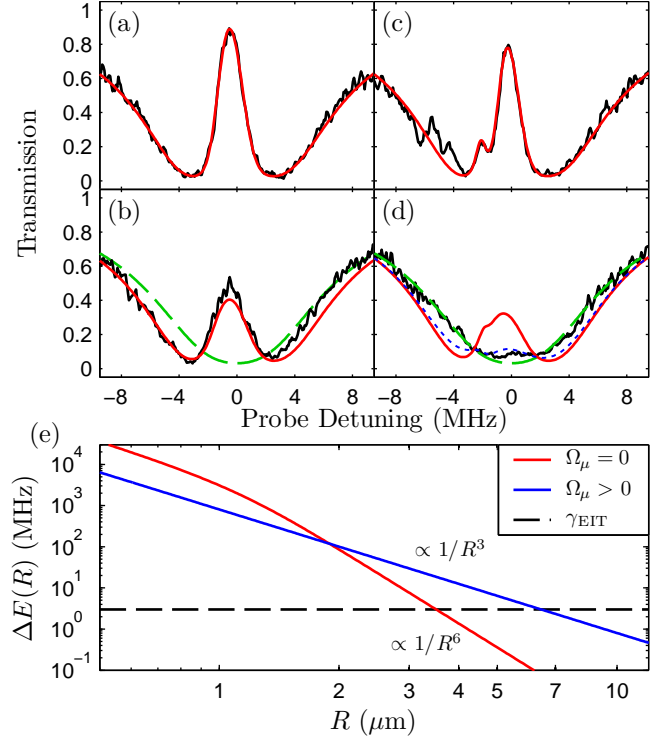}
\caption[]{Enhanced suppression of EIT for weak microwave dressing. Experimental spectra (solid black curves) and theoretical modelling for EIT in the weak probe regime $\Omega_{\rm p}/2\pi$ = 80 kHz ((a)and (c)) and strong probe regime $\Omega_{\rm p}/2\pi$ = 1.9 MHz ((b) and (d)). (a) and (b): no microwave dressing. With weak microwave dressing, (c) and (d), there is almost complete suppression of the EIT signal in the strong probe regime. 
Red solid curves: 
transmission calculated as explained in Section 5.
Blue dotted curve: the same as the red solid curve but assuming the $1/R^3$ dipole-dipole 
interaction.
Green dashed curves: transmission calculated assuming complete blockade of Rydberg excitation.
Plot (e) shows the energy shift arising from the interaction between
Rydberg states as a function of the interatomic separation $R$: the microwaves create a long-range $1/R^3$ interaction, increasing the blockade radius. (The shift is expressed as a frequency.)\label{fig5}}
\end{center}
\end{figure}

\section{Conclusions and Outlook}
In conclusion, we have demonstrated microwave dressing of electromagnetically induced transparency involving highly excited Rydberg states. The microwave field splits the EIT peak resulting in independent control of the absorptive and dispersive properties of the medium. Consequently a microwave field could be used to control the interaction time between Rydberg polaritons. In addition we demonstrate that microwave dressing leads to enhanced interactions due to an effective increase in the blockade radius.
In particular, we present evidence for a 
$1/R^3$ energy shift between Rydberg
states resonantly coupled by the microwave field and the ensuing breakdown of the pair-wise interaction approximation. Such microwave tuning of the non-linear optical response of the blockaded ensemble could be useful in the realisation of single photon phase gates \cite{fried05}. The microwave dressing could also prove useful for detection of atoms in states that can be brought in to F\"orster resonances with relevant Rydberg states.

\appendix
\section{Relaxation Matrix $\mathcal{L}_d(\rho)$}\label{sec:L}
The effect of the finite laser linewidth is to cause a dephasing of the off-diagonal coherence terms in the density matrix \cite{gea95}. The resulting dephasing matrix is
\begin{equation}
\mathcal{L}_d(\rho)=-\displaystyle\sum_{i,j} \gamma_{i,j}\rho_{i,j}\ket{i}\bra{j},
\end{equation}
where the laser-induced dephasing rates $\gamma_{i,j}$ are obtained from summing over the linewidth of all fields coupling $\ket{i}$ to $\ket{j}$. Replacing the terms $\gp+\gc\rightarrow\gr$ for the arguments given above, the total dephasing rates are given by
\begin{equation}
{\scriptsize\fl \gamma\!=\!\!\left(\!\!\!\begin{array}{cccccccccc}
0&\gp&\gr&\gr\!+\!\gamma_{{\rm Ry}}&\gr\!+\!\gamma_{{\rm Ry}}&\gr\!+\!\gamma_{{\rm Ry}}&\gr\!+\!\gamma_{{\rm Ry}}&\gr\!+\!\gamma_{{\rm Ry}}&\gr\!+\!\gamma_{{\rm Ry}}&\gr\!+\!\gamma_{{\rm Ry}} \\
\gp&0&\gc&\gc\!+\!\gamma_{{\rm Ry}}&\gc\!+\!\gamma_{{\rm Ry}}&\gc\!+\!\gamma_{{\rm Ry}}&\gc\!+\!\gamma_{{\rm Ry}}&\gc\!+\!\gamma_{{\rm Ry}}&\gc\!+\!\gamma_{{\rm Ry}}&\gc\!+\!\gamma_{{\rm Ry}} \\
\gr&\gc&0&\gamma_{{\rm Ry}}&\gamma_{{\rm Ry}}&\gamma_{{\rm Ry}}&\gamma_{{\rm Ry}}&\gamma_{{\rm Ry}}&\gamma_{{\rm Ry}}&\gamma_{{\rm Ry}}\\
\gr\!+\!\gamma_{{\rm Ry}}&\gc\!+\!\gamma_{{\rm Ry}}&\gamma_{{\rm Ry}}&0&0&\gamma_{{\rm Ry}}&\gamma_{{\rm Ry}}&\gamma_{{\rm Ry}}&\gamma_{{\rm Ry}}&\gamma_{{\rm Ry}} \\
\gr\!+\!\gamma_{{\rm Ry}}&\gc\!+\!\gamma_{{\rm Ry}}&\gamma_{{\rm Ry}}&0&0&\gamma_{{\rm Ry}}&\gamma_{{\rm Ry}}&\gamma_{{\rm Ry}}&\gamma_{{\rm Ry}}&\gamma_{{\rm Ry}} \\
\gr\!+\!\gamma_{{\rm Ry}}&\gc\!+\!\gamma_{{\rm Ry}}&\gamma_{{\rm Ry}}&\gamma_{{\rm Ry}}&\gamma_{{\rm Ry}}&0&0&\gamma_{{\rm Ry}}&\gamma_{{\rm Ry}}&\gamma_{{\rm Ry}} \\
\gr\!+\!\gamma_{{\rm Ry}}&\gc\!+\!\gamma_{{\rm Ry}}&\gamma_{{\rm Ry}}&\gamma_{{\rm Ry}}&\gamma_{{\rm Ry}}&0&0&\gamma_{{\rm Ry}}&\gamma_{{\rm Ry}}&\gamma_{{\rm Ry}} \\
\gr\!+\!\gamma_{{\rm Ry}}&\gc\!+\!\gamma_{{\rm Ry}}&\gamma_{{\rm Ry}}&\gamma_{{\rm Ry}}&\gamma_{{\rm Ry}}&\gamma_{{\rm Ry}}&\gamma_{{\rm Ry}}&0&0&\gamma_{{\rm Ry}} \\
\gr\!+\!\gamma_{{\rm Ry}}&\gc\!+\!\gamma_{{\rm Ry}}&\gamma_{{\rm Ry}}&\gamma_{{\rm Ry}}&\gamma_{{\rm Ry}}&\gamma_{{\rm Ry}}&\gamma_{{\rm Ry}}&0&0&\gamma_{{\rm Ry}} \\
\gr\!+\!\gamma_{{\rm Ry}}&\gc\!+\!\gamma_{{\rm Ry}}&\!\gamma_{{\rm Ry}}&\!\gamma_{{\rm Ry}}&\!\gamma_{{\rm Ry}}&\!\gamma_{{\rm Ry}}&\!\gamma_{{\rm Ry}}&\!\gamma_{{\rm Ry}}&\!\gamma_{{\rm Ry}}&0
\end{array}\!\!\!\right) }.
\end{equation}

\section*{Acknowledgments}
We thank the EPSRC, Durham University and the DPST Programme of the
Thai Government for financial support.

\end{document}